\documentclass[prd,aps,floats,preprint,preprintnumbers]{revtex4}
\usepackage{amsmath,amssymb}
\usepackage{graphicx}
\usepackage{graphics}
\usepackage{epsfig}
\usepackage{verbatim}
\usepackage{html,makeidx}
\makeindex

\numberwithin{equation}{section}

\newcommand {\eqv}{\equiv}

\newcommand {\ora}{\overrightarrow}
\newcommand {\ola}{\overleftarrow}

\newcommand {\td}{\tilde}
\newcommand {\dg}{\dagger}
\newcommand {\ra}{\rightarrow}
\newcommand {\Ra}{\Rightarrow}

\newcommand {\vep}{\varepsilon}
\newcommand {\vphi}{\varphi}
\newcommand {\del}{\partial}

\newcommand {\ld}{\lambda}

\newcommand {\al}{\alpha}
\newcommand {\bi}{\beta}

\newcommand {\bfr}{\begin{flushright}}
\newcommand {\efr}{\end{flushright}}
\newcommand {\bfl}{\begin{flushleft}}
\newcommand {\efl}{\end{flushleft}}

\newcommand {\nn} {\nonumber}

\newcommand {\txt}{\textrm}
\newcommand {\bd}{
\begin{document}}
\newcommand {\ed}{\end{document}}

\newcommand {\be}{\begin{equation}}
\newcommand {\ee}{\end{equation}}
\newcommand {\bea}{\begin{eqnarray}}
\newcommand {\eea}{\end{eqnarray}}
\newcommand {\ba}{\begin{array}}
\newcommand {\ea}{\end{array}}

\newcommand{\bbib}{\begin{thebibliography}{99}}
\newcommand{\ebib}{\end{thebibliography}}
\newcommand {\bab}{\begin{abstract}}
\newcommand {\eab}{\end{abstract}}

\newcommand {\bc}{\begin{center}}
\newcommand {\ec}{\end{center}}
\newcommand {\bit}{\begin{itemize}}
\newcommand {\eit}{\end{itemize}}
\newcommand {\ul}{\underline}
\newcommand {\txtc}{\textcolor}

\newcommand {\ad}{\textrm{ad}}
\newcommand {\Ad}{\textrm{Ad}}
\def\A{{\cal A}}\def\B{{\cal B}}\def\C{{\cal C}}\def\D{{\cal D}}\def\E{{\cal E}}\def\F{{\cal F}}\def\G{{\cal G}}\def\H{{\cal H}}\def\I{{\cal I}}
\def\J{{\cal J}}\def\K{{\cal K}}\def\L{{\cal L}}\def\M{{\cal M}}\def\N{{\cal N}}\def\O{{\cal O}}\def\P{{\cal P}}\def\Q{{\cal Q}}\def\R{{\cal R}}
\def\S{{\cal S}}\def\T{{\cal T}}\def\U{{\cal U}}\def\V{{\cal V}}\def\W{{\cal W}}\def\X{{\cal X}}\def\Y{{\cal Y}}\def\Z{{\cal Z}}

\def\xbar{\bar{x}}\def\ybar{\bar{y}}\def\zbar{\bar{z}}\def\kbar{\bar{k}}\def\pbar{\bar{p}}


\bd

\small
\preprint{SU-4252-891 \vspace{1cm}} \setlength{\unitlength}{1mm}
\title{Curving Space, Effective Gravity and Simultaneous Measurements
\vspace{0.5cm}}
\author{E. Akofor}\thanks{eakofor@phy.syr.edu}\affiliation{Department of Physics, Syracuse University, Syracuse, NY
13244-1130, USA \vspace{0cm}}

\begin{abstract}
\vspace{0.5cm}
We present a way of understanding the curvature of space-time, the basic philosophy being that the (linear) geometry of any space is determined by the (linear) functionals on the algebra(s) of any fields defined on the space.

It is known that quantum states or hypothetical measurements on a quantum system may be regarded as unit or normalized positive linear functionals on a $C^\ast$-algebra which contains the variable events (or simply ``variables'' to be brief) that describe the dynamics of the quantum system.
We consider linear projection geometry and differential calculus on a $C^\ast$-algebra $\mathcal{A}$ by means of the positive linear functionals $\mathcal{A}^\ast$ on $\mathcal{A}$. The analysis is done with the help of linear projections based on a version of the Cauchy-Schwarz inequality. We mention the relation between uncertainties or errors involved in the simultaneous measurement of more than one statistical or quantum variable. We also attempt to obtain Einstein's theory of the gravitational field by analytic continuation.

\end{abstract}
\maketitle

\section{Introduction}\label{intro}

Our senses and measurement instruments may be sensitive only to projections of real events onto our own frame(s) of reference. It is however possible to learn more about the actual situations or events if we know how these events project onto our reference frames. This fact underlies the importance of projections in our context.

We will represent variable events by elements of a $C^\ast$-algebra $\mathcal{A}$, then in a particular measurement or ``state'' $\phi\in\mathcal{A}_+^\ast$ (the set of positive linear functionals on $\mathcal{A}$), an observer can project a particular event $A$ onto the ''hyper-plane'' $Span\{B_i\}$ spanned by a chosen collection of reference events $\{B_i\in \A,~i=1,2,...,n\}$ ($Span\{B_i\}=\{B=\al_i B_i,~~\al\in \mathbb{C}^p\}$), and with the help of  $\phi$, determine the projection coefficients of $A$ with respect to $\{B_i\}$. These projection coefficients measure  correlations or interdependencies between the event $A$ and the reference event set $\{B_i\}$. An event will be taken to be any descriptive label that is partially quantitative and partially qualitative (for example, a special complex matrix).

In this section, we define the $C^\ast$-algebra appropriate for our purpose. In the following section \ref{preliminaries}, we review the calculus of hypersurfaces (or submanifolds) embedded in  $\mathbb{R}^n$ \cite{frankel} . In section \ref{projections}, with the help of the Cauchy-Schwarz inequality, we define projections which we then use to carry out differential calculus on a $C^\ast$-algebra in section \ref{calculus}.

A $C^\ast$-algebra $\mathcal{A}$ will be taken to be an associative algebra over the field of complex numbers $\mathbb{C}$ which is closed under an operation $^\ast$ (that is, $a^\ast\in \mathcal{A}~~\forall a\in \mathcal{A}$) with the following properties.
\bea
&&a^\ast{}^\ast=a,~~(ab)^\ast=b^\ast a^\ast,~~(a+b)^\ast=a^\ast+b^\ast,\nn\\
&&\al^\ast=\bar{\al}~~~~\forall~a,b\in \mathcal{A},~~\al\in \mathbb{C},
\eea
where $\bar{\al}$ denotes the complex conjugate of $\al$.
That is,
\bea
&&\A=(A,\mathbb{C},+,\star,^\ast)\nn\\
&&~~~~=\{a\in A;~a(bc)=(ab)c,~(a+b),(ab:=a\star b),a^\ast,\ld a\in A~~\forall a,b\in A~\&~\forall \ld\in \mathbb{C}\}.
\eea

A linear functional $\phi\in \mathcal{A}^\ast$  on $\mathcal{A}$ has the following properties
\bea
&&\phi: \mathcal{A}\ra \mathbb{C},~a\mapsto \phi(a),\nn\\
&&\phi(a^\ast)=\overline{\phi(a)},\nn\\
&&\phi(a+b)=\phi(a)+\phi(b),\nn\\
&&\phi(\al a)=\al \phi(a)~~~~\forall~a,b\in \mathcal{A},~~\al\in \mathbb{C}.
\eea
The set of states or (normalized) positive linear functionals is given by
\bea
\mathcal{A}_+^\ast=\{\phi\in \mathcal{A}^\ast,~\phi(1_\mathcal{A})=1,~\phi(a^\ast a)\in \mathbb{R}^+\subset \mathbb{C}\},
\eea
where $1_\A$ is the multiplicative identity in $\A$. The positivity of $\phi$ allows one to define probability density functions.

\section{Preliminaries: Hypersurfaces in $\mathbb{R}^n$}\label{preliminaries}
The linear geometry of a $p$-dimensional hypersurface $\U$ in $\mathbb{R}^n$  is determined by the linear functionals on the algebra of $\mathbb{R}^n$-valued smooth functions on $\U$. The validity of the previous statement depends on facts similar to that the usual dot-product on $\mathbb{R}^n$ can be realized with a linear functional, the trace ${1\over n}\txt{Tr}$ over the algebra of real $n\times n$ diagonal matrices $\txt{diag}(\mathbb{R}^n)$ which is isomorphic to $\mathbb{R}^n$. See section (\ref{ssec:Project-linear-depende}).

 While in a p-dimensional subspace~ $\V=\{\vec{x}(u),~u=(u^\al),~\al=1,2,...,p\}$~ of~ $\mathbb{R}^n=\{x\}$, we may regard the coordinate functions $\vec{x}$ as vector fields $\vec{x}:\U\ra \V\subset \mathbb{R}^n,~~u\ra \vec{x}(u)$ defined on the parameter space  ~$\U=\{u\}$ of $\V=\{\vec{x}(u)\}$~ with values in $\mathbb{R}^n$.

That is, the curvature of $\U\simeq \{u\}$ is determined by the fields~ $\vec{x}:\U\ra \V\simeq \{\vec{x}(u)\}\subset \mathbb{R}^n\simeq \{x\}$ ~ on $\U$. As usual, the induced metric $G$ in the local coordinates $u$ on the hypersurface $\U$ is obtained by the dot-product of the partial derivatives of the vector fields $\vec{x}$. That is,
\bea
G=(g_{\al\beta})=(\vec{x}_\al\cdot\vec{x}_\beta),~~~~\vec{x}_\al=\del_\al\vec{x}={\del\vec{x}\over\del u^\al}.
\eea
The space of tangent vectors  $T_u(\U)$ at $u$ is spanned by the set $\{\vec{x}_\al,~~\al=1,2,...,p\}$. That is,
\bea
T_u(\U)=\{\vec{v}=\td{v}^\al \vec{x}_\al\}.
\eea
A ``standard'' way to displace (or infinitesimally transport) a vector $\vec{v}$ along $\U$ while maintaining tangency to $\U$ is to use the covariant (or projected) derivative defined as follows.

\bea
&&\nabla: T_u(\U)\ra T_u(\U),~\vec{v}\ra \nabla\vec{v}=P\cdot d\vec{v}=\vec{x}_\al g^{\al\beta}\vec{x}_\beta\cdot d\vec{v}.\nn\\
&&g^{\al\beta}=g^{-1}_{\al\beta}.
\eea
That is, $\nabla$ is the projection through the projector $P=g^{\al\beta}\vec{x}_\al\otimes \vec{x}_\beta\eqv\vec{x}_\al g^{\al\beta}\vec{x}_\beta\eqv (x^i_\al g^{\al\beta}x^j_\beta) $ of the differential $d$ along the tangent plane at $u\in \U$.

Since the metric here is symmetric, $g_{\al\beta}=g_{\beta\al}$, one can then check that
\bea
&&\nabla\vec{v}=\vec{x}_\al(d\td{v}^\al+\Gamma^\al{}_{\beta}\td{v}^\beta)\eqv \vec{x}_\al D\td{v}^\al,\nn\\
&& \Gamma^\al{}_\rho=g^{\al\beta}\vec{x}_\beta\cdot d\vec{x}_\rho={1\over 2}~ g^{\al\beta}~(\del_{\rho}g_{\beta\delta}-\del_\beta g_{\rho\delta}+\del_{\delta} g_{\rho\beta})du^{\delta},
\eea
where $D$ is the representation of $\nabla$ on the vectors given in component form. One can take higher derivatives and the second derivative $\nabla^2$ will contain information about the curvature of $\U$. The curvature $2$-form of the hypersurface $\U$ can then be expressed as
\bea
R^\al{}_\beta=\hat{d}\Gamma^\al{}_\beta+\Gamma^\al{}_\rho\wedge\Gamma^\rho{}_\beta~~\txt{or simply}~~{d}\Gamma^\al{}_\beta+\Gamma^\al{}_\rho\Gamma^\rho{}_\beta ,
\eea
where $\hat{d}$ is the exterior derivative (an anti-symmetrized tensorial derivative, section \ref{tensoring-derivative}) and $\wedge$ denotes the exterior product (an anti-symmetrized tensor product).
For more information, see references such as \cite{frankel},\cite{majid},\cite{nakahara},\cite{bertlemann}.

We will talk about tensors in section
 {\ref{calculus}}.

\section{Linear Projection Geometry}\label{projections}
This section builds on the definitions made in the introduction (section \ref{intro}) regarding $C^\ast$-algebras.

\subsection{Cauchy-Schwarz inequality}\label{CSI}
Consider any $A\in \mathcal{A}$,~ a collection $\{B_i\in \mathcal{A},~~i=1,2,...,p\}$ and $\phi\in \mathcal{A}^\ast_+$. Also define the function ~$f:\mathbb{R}^n\ra \mathbb{R}^+,~\ld\ra f(\ld)$,

\bea
&&f(\ld)=\phi((A+\ld_i B_i)^\ast(A+\ld_i B_i))=\phi(A^\ast A)+\ld_i\phi(A^\ast B_i+B_i^\ast A )+\ld_i\ld_j\phi(B_i^\ast B_j)\nn\\
&&~~~~\eqv \phi(A^\ast A)+2\ld_i N_i+\ld_i\ld_j M_{ij}\geq 0,\\
&&N_i={1\over 2}\phi(A^\ast B_i+B_i^\ast A ),~~M_{ij}={1\over 2}\phi(B_i^\ast B_j+ B_j^\ast B_i).
\eea
The value of $f$ at its extreme point gives the Cauchy-Schwarz inequality. That is,
\bea
\label{lin-indep}&&{\del\over\del \ld_i}f(\ld')=0~~\iff~~ \ld'_i=-M^{-1}_{ij}N_j,~~\det (M_{ij})> 0,\\
\label{cauchy-inequal}&&~~\Ra~~f(\ld')=\phi(A^\ast A)-N_iM^{-1}_{ij}N_j\geq 0.
\eea

If we write~ $B_I=(A,B_i)\eqv (B_0,B_i),~~M_{IJ}={1\over 2} \phi(B_I^\ast B_J+ B_J^\ast B_I),~~I,J=0,1,...,p$~ then the Cauchy-Schwarz inequality (\ref{cauchy-inequal}) becomes
\bea
\label{cauchy-inequal-2}&&f(\ld')={\det (M_{IJ})\over \det (M_{ij})}\geq 0~~\Ra~~ \det (M_{IJ})\geq 0.
\eea

\subsection{Linear projections and linear dependence}\label{ssec:Project-linear-depende}
The positive determinant condition ~$\det (M_{ij})> 0$~ of (\ref{lin-indep}) is identified as the requirement for the linear independence of the set $\{B_i\}$ in the state $\phi$.
Any collection of $p$ elements $\{B_i\in \mathcal{A},~~i=1,2,...,p\}$ of $\mathcal{A}$ is linearly independent in the state $\phi\in \mathcal{A}^\ast_+$ if
\bea
&&\det M_{p\times p}>0,~~ M_{p\times p}=(M_{ij}),~~M_{ij}={1\over 2}\phi(B^\ast_i B_j+B^\ast_j B_i).
\eea
We also identify $N_iM^{-1}_{ij}N_j$ of (\ref{cauchy-inequal})  as the projection $\txt{Proj}_{\{B_i\}}(A)\eqv A|_{\{B_i\}}$ of $A$ onto the hyperplane $\txt{Span}\{B_i\}$ spanned by $\{B_i\}$.
\bea
&&\txt{Proj}_{\{B_i\}}(A)~^2=N_iM^{-1}_{ij}N_j,~~N_i={1\over 2}\phi(A^\ast B_i+B_i^\ast A ).
\eea

Equality in (\ref{cauchy-inequal}) holds when $A+\ld'_i B_i=0$. That is, equality holds when $A\in \txt{Span}\{B_i\}$ in the state $\phi$.
\bea
A=-\ld'_iB_i=B_iM^{-1}_{ij}N_j.
\eea

As an example (see section \ref{CSI} for notation), suppose $\mathcal{A}=\{a=\txt{diag}(\vec{a}),~\vec{a}=(a^\al)\in \mathbb{R}^n\}$ is the set of real diagonal $n\times n$ matrices which is isomorphic to $\mathbb{R}^n$ and~ $\phi(a)={1\over n}\txt{Tr}~a,~~a^\ast=a^\dagger$. Then
\bea
&&\det M_{IJ}={1\over n^{p+1}}\det (\vec{b}_I\cdot\vec{b}_J)={1\over n^{p+1}}~f_{\al_1...\al_{n-p-1}}f^{\al_1...\al_{n-p-1}},\nn\\
&&f_{\al_1...\al_{n-p-1}}={1\over \sqrt{(n-p-1)!}}\vep_{\al_1...\al_{n-p-1}\beta_0\beta_1...\beta_p}a^{\beta_0}b_1^{\beta_1}...b_p^{\beta_p},
\eea
gives the volume of the parallelepipede formed by the $p+1$ vectors $\vec{b}_I=\vec{a},\vec{b}_1,...,\vec{b}_p\in \mathbb{R}^n$,~ where~ $\vep_{\al_1...\al_{n-p-1}\beta_0\beta_1...\beta_p}$~ is the Levi-Civita symbol.

For $p=2$, let $\vec{b}_1=\vec{b},~~\vec{b}_2=\vec{c}$~~ with ~$\vec{b},\vec{c}\in \mathbb{R}^n$ of course. Then
\bea
&& \nn\\
&& a|_{(\vec{b},\vec{c})}~^2=\left(
                     \begin{array}{cc}
                       \vec{a}\cdot\vec{b} & \vec{a}\cdot\vec{c} \\
                     \end{array}
                   \right)\left(
                            \begin{array}{cc}
                              \vec{b}^2 & \vec{b}\cdot\vec{c} \\
                              \vec{b}\cdot\vec{c} & \vec{c}^2 \\
                            \end{array}
                          \right)^{-1}\left(
                                        \begin{array}{c}
                                          \vec{a}\cdot\vec{b} \\
                                          \vec{a}\cdot\vec{c} \\
                                        \end{array}
                                      \right)={(\vec{a}\cdot\vec{b})^2\vec{c}^2-2~\vec{a}\cdot\vec{b}~~\vec{a}\cdot\vec{c}~~\vec{b}\cdot\vec{c}+(\vec{a}\cdot\vec{c})^2\vec{b}^2\over \vec{b}^2\vec{c}^2-(\vec{b}\cdot\vec{c})^2}\nn\\
&&~~~~=(~{\vec{a}\cdot\vec{c}~~\vec{b}-\vec{a}\cdot\vec{b}~~\vec{c} \over \sqrt{\vec{b}^2\vec{c}^2-(\vec{b}\cdot\vec{c})^2}}~)^2,\nn\\
\label{CSI-angles}&&\vec{a}^2\geq a|_{(\vec{b},\vec{c})}~^2~~\Ra~~ \cos^2\theta_1+\cos^2\theta_2+\cos^2\theta\leq 1+2\cos\theta_1\cos\theta_2\cos\theta,
\eea
where~ $\theta_1=\angle \vec{a}\vec{b},~\theta_2=\angle \vec{a}\vec{c},~\theta=\angle \vec{b}\vec{c}$. For a visualization, one can directly verify the inequality (\ref{CSI-angles}) for the case of~ $\vec{a},\vec{b},\vec{c}\in\mathbb{R}^3$ where
the closed tetrahedral surface
\bea
S=\{(x,y,z)\in [-1,1]^3,~x^2+y^2+z^2=1+2xyz\}
\eea
generates the convex set
\bea
C=\{(x,y,z)\in [-1,1]^3,~x^2+y^2+z^2\leq 1+2xyz\}.
\eea

As another example, given an $n\times n$ hermitian matrix $A^\dagger=A$, one can seek a linear dependence relationship among the powers $B_i=A^i=AA^{i-1},~~i=0,1,2,...,m$~~ of ~$A$~ for some positive integer $m$ (which can also be infinite). Then one finds ~~$\mathbb{I}+\al_i A^i=0,~~\al_i=-(\txt{Tr}(A^iA^j))^{-1}\txt{Tr}A^j$, which may be related to the characteristic polynomial(s) of the matrix $A$. Indeed if such a relation exists, it will be satisfied by the eigenvalues of $A$ since one can write $A=PA_\ld P^{-1},~~A_\ld=\txt{diag}
(\ld_1,\ld_2,...,\ld_n)$. One may expect only a special class of hermitian matrices to satisfy such a relation.
Alternatively, the components of each eigenvector $\al=(\al_i)$ of the ``power'' matrix ~$M_{i~ab}=(A^i)_{ab}\eqv M_{ij},~~j=n(a-1)+b,~~a,b=1,2...n,~i,j=1,2...n^2$~ form the coefficients of a characteristic polynomial of $A$.

\subsection{Simultaneous Measurements and Uncertainties}\label{SME}
Practical measurements are often simultaneous measurements involving more than one observable. In this section $\ref{SME}$ we will denote the state $\phi$ by $\langle~\rangle$.

The fluctuation $\delta A=A-\langle A\rangle $ in an operator $A$ naturally projects onto the planes defined by sets of operators, including the plane of fluctuations $\{\delta B_i=B_i-\langle B_i\rangle\}$, just as the operator $A$ does. This introduces uncertainty bounds in measurements since  measurement uncertainty estimates of the type $\Delta A{}^2= \langle \delta A^\dg \delta A\rangle$ and $\{\Delta B_i{}^2= \langle \delta B_i^\dg \delta B_i\rangle\}$ are related by the inequality (\ref{cauchy-inequal}),~(\ref{cauchy-inequal-2}).

\subsubsection{Momentum and position in quantum mechanics}

Consider the momentum and position operators $p_i=p_i(t)$ and $x_i=x_i(t)$, then
\bea
\label{canonical}&&[p_i,x_j]=-i\hbar\delta_{ij},~~[p_i,p_j]=0,~~[x_i,x_j]=0,\\
\label{canonical2}&&\langle p^2_i\rangle \langle x^2_j\rangle\geq {\hbar^2\over 4}\delta_{ij},
\eea
if only two operators are measured simultaneously.

This means that the momentum and position in a particular direction $i=j$ are not independent and so their simultaneous measurements interfere with one another up to an uncertainty determined by the reduced Plank constant $\hbar$. On the other hand, the simultaneous measurements of position and momentum projections in two orthogonal directions $i\neq j$ do not affect one another.

If the observables in more than two directions are considered simultaneously, then one would need to use the inequality (\ref{cauchy-inequal}),~(\ref{cauchy-inequal-2}) in full.

\subsubsection{Energy Bound}
Let $A=H(t),~~H^\dg=H$ be the Hamiltonian for a physical system whose space of observables includes the set $\{B_i(t)\}$ of hermitian/antihermitian operators. Then
\bea
\label{energy}&&\langle H^2\rangle\geq {\hbar^2\over 4}\langle\dot{B}_i\rangle M^{-1}_{ij}\langle\dot{B}_j\rangle,~~M_{ij}={1\over 2}\langle \{B_i,B_j\}\rangle={1\over 2}\langle B_iB_j+B_jB_i \rangle,\nn\\
&&[H,B_i]=-i\hbar({dB_i\over dt}-{\del B_i\over \del t})\eqv -i\hbar\dot{B}_i,
\eea
 where $h(B_i)~^2={\hbar^2\over 4}\langle\dot{B}_i\rangle M^{-1}_{ij}\langle\dot{B}_j\rangle$ may be regarded as the lower bound for a canonical ``excitation/thermal'' energy associated with the system of observables $\{B_i\}$ during a simultaneous measurement in the state $\phi=\langle~\rangle$. The larger $h(B_i)$ the more excited the system $\{B_i\}$ will be in the state $\phi=\langle~\rangle$ and vice versa. If energy is required or involved in the measurement of $\{B_i\}$ and we try to \emph{simultaneously} measure $\{B_i\}$ to arbitrarily high accuracies, then we will need perhaps unattainably high energy in order to do so if $\langle \dot{B}_i\rangle\neq 0$. This is due to the following inequality for uncertainties.
 \bea
 &&\Delta H~^2=\langle \delta H~^2\rangle\geq {\hbar^2\over 4}\langle\dot{B}_i\rangle m^{-1}_{ij}\langle\dot{B}_j\rangle,~~m_{ij}={1\over 2}\langle \{\delta B_i,\delta B_j\}\rangle={1\over 2}\langle \delta B_i\delta B_j+\delta B_j\delta B_i \rangle.\nn\\
 \eea

\subsubsection{Noncommutative Space-time}
Studies in quantum gravity suggest that there needs to be a finite lower limit to the separation between two events in space-time that can allow for the two events to be distinguishable by experiment.
To introduce this minimum length in a canonical way (\ref{canonical}),(\ref{canonical2}),(\ref{energy}) one encounters non-commutativity among position operators $q^\mu$ in different directions $\mu=0,1,2...,n$~~ \cite{KMM1}\cite{KMM2}.
In the case of a Moyal space-time, $[x_\mu,x_\nu]=i\theta_{\mu\nu}=-i\theta_{\nu\mu}=constant$, if only two operators are measured simultaneously, then
\bea
&&\langle x^2_\mu\rangle \langle x^2_\nu\rangle\geq {1\over 4}\theta_{\mu\nu}{}^2.
\eea
Again, if the observables in more than two directions are considered simultaneously, then one would need to use the inequality (\ref{cauchy-inequal}),~(\ref{cauchy-inequal-2}) in full.

It is worthwhile to mention that the extended dot-product of the following section should not be confused with the usual dot-product in $\mathbb{R}^n$ used in the previous sections. From here onward any dot-product will be the extended dot-product unless specified otherwise.

\subsection{Decomposition and basis}
While in the state $\phi$, let's define a dot-product $\cdot$ (to simplify notation) and the kernel $\ker a$ of $a\in \mathcal{A}$ as follows.
\bea
&&\cdot:\mathcal{A}\times \mathcal{A}\ra \mathbb{C},~~(a,b)\mapsto \cdot(a,b)=a\cdot b=b\cdot a~:=~{1\over 2}~\phi(a^\ast b+b^\ast a).\nn\\
&&\ker a=\{c\in \mathcal{A},~~a\cdot c=c\cdot a=0\}.
\eea

Then we can decompose $a\in \mathcal{A}$ in the state $\phi$ in terms of $\{b_i\}$ thus
\bea
&&a=a^{\perp}+a^{\parallel}\eqv a|_{\ker\{b_i\}}+a|_{\{b_i\}}\eqv \txt{Proj}_{\ker\{b_i\}}(a)+\txt{Proj}_{\{b_i\}}(a),
\eea
where
\bea
&&a^{\parallel}= a|_{\{b_i\}}=\txt{Proj}_{\{b_i\}}(a) =b_i~(b_i\cdot b_j)^{-1}~b_j\cdot a,\nn\\
&&a^{\perp}=a|_{\ker\{b_i\}}=\txt{Proj}_{\ker\{b_i\}}(a) =\td{b}_\mu~(\td{b}_\mu\cdot \td{b}_\nu)^{-1}~\td{b}_\nu\cdot a,\nn\\
&&\ker\{b_i\}=\ker b_1 \cap \ker b_2\cap...\cap\ker b_p=\txt{Span}\{\td{b}_\mu\} ,\nn\\
&&\det (\td{b}_\mu\cdot \td{b}_\nu)>0.
\eea
That is, $\{\td{b}_\mu\}$  is any linearly independent set in $\ker\{b_i\}$ that spans (is a basis for) $\ker\{b_i\}$.
The linearly independent set $\{b_i\}$ forms a basis for $\mathcal{A}$ in the state $\phi$ if~ $\ker\{b_i\}=\{0_\mathcal{A}\}$.

If we denote the projector onto the $\{b_i\}$ plane by $P$, then the mirror image (or reflection) $a_{rfl}$ of $a$ about the $\{b_i\}$ plane is
\bea
&&a_{rfl}=-a^{\perp}+a^\parallel=\{-(\mathbb{I}-P)+P\}\cdot a=(2P-\mathbb{I})\cdot a~,\nn\\
&&(2P-\mathbb{I})\cdot (2P-\mathbb{I})=\mathbb{I}~~\Ra~~a_{rfl}\cdot a_{rfl}=a\cdot a
\eea

\subsection{Orthogonalization}\label{Ortho}
Consider the basis set $B_k=\{b_i,~~i=1,2...,k\}=\{b_1,b_2,...,b_k\}$ and similarly label the following sequence of $k$ subsets as $B_i=\{b_1,b_2,...,b_i\},~~i=1,2,...,k$. Then as before,\\ $P_i=\txt{Proj}_{B_{i-1}}=b_\al(b_\al\cdot b_\beta)^{-1}b_\beta,~~\al,\beta=1,2,...,i-1$~~ is the projector unto the ``$(i-1)$-plane'' spanned by the set $B_i$.
The sequence of projectors $(P_i)$ has the following basic properties:
\bea
&& P_i\cdot P_j=\left\{
                  \begin{array}{ll}
                    P_i, &~~ i \leqslant j \\
                    P_j, &~~ i \geqslant j
                  \end{array}
                \right\},~
  \nn\\
&& P_i\cdot b_j=b_j~~\txt{for}~~i>j.~~~~
\eea

One can therefore check that the new set of objects $O_k=\{o_i,~i=1,2,3...,k\}$ is orthogonal ($o_i\cdot o_j=0$~ for $i\neq j$), where
\bea
&&o_i=(\mathbb{I}-P_i)\cdot b_i,~~ i=1,2,...k.\nn\\
&&P_1=0,\nn\\
&&o_1=(\mathbb{I}-P_1)\cdot b_1=b_1,\nn\\
&&o_2=(\mathbb{I}-P_2)\cdot b_2=(\mathbb{I}-b_1 {1\over b_1\cdot b_1} b_1)\cdot b_2=b_2-b_1~{b_1\cdot b_2\over b_1\cdot b_1},\nn\\
&&...\nn\\
&&o_k=(\mathbb{I}-P_k)\cdot b_k.\nn\\
&&o_i\cdot o_j=b_i\cdot(\mathbb{I}-P_i)\cdot (\mathbb{I}-P_j)\cdot b_j.
\eea
We can then obtain the orthonormal set $\hat{O}_k=\{\hat{o}_i,~~i=1,2,...,k,~~\hat{o}_r\cdot\hat{o}_s=\delta_{rs}\}$
\bea
\hat{o}_i={o_i\over \sqrt{o_i\cdot o_i}}={(\mathbb{I}-P_i)\cdot b_i\over \sqrt{b_i\cdot(\mathbb{I}-P_i)\cdot b_i}} .
\eea
See appendix \ref{orthon-frame} for an application of orthonormalization.

\subsection{Non-linear inner product on $\A^n$}
For each state $\phi$ one has on $\A^n=\A\times \A^{n-1}=\{A=(a_i)=(a_1,a_2,...,a_n),~~a_1,...a_n\in\A\}$,  the non-linear inner product
\bea
&&( A,B){}^n=\det (a_i\cdot b_j),~~a_i\cdot b_j={1\over 2}\phi(a^\ast_i b_j+b_i a^\ast_j),~~A,B\in \A^n,\nn\\
&&( A,A)\geq 0.\nn
\eea

\section{Differential Calculus: Hypersurfaces in $\A$}\label{calculus}
The (linear) geometry of any space is determined by the (linear) functionals on the algebra of fields on the space. Here we replace the $\mathbb{R}^n$ of section \ref{preliminaries} with a $C^\ast$-algebra $\A$.

\subsection{Projected derivative}
Consider a parametrized proper subset (with parameters $x\in\mathbb{R}^n$) ~$\V\simeq\{b(x)\}\subset \mathcal{A}$ ~of the~ $C^\ast$-algebra $\mathcal{A}=\{a\}$~, where $\simeq$ means ``locally isomorphic to''. Here, $x\in \mathbb{R}^n$ may be thought of as local coordinates on a parameter space $\U\simeq \mathbb{R}^n$ of $\V$. $\V$ may also be thought of as the value space or image of a map ~$b:\U\ra \A$. Consider the partial derivative- or ``coordinate''-basis $\{b_\al={\del b(x)\over \del x^\al}=\del_\al b(x)\}$ for the tangent space $T_x(\U)$ to $\U$ at the point $x$.  That is, $T_x(\U)=\txt{Span}\{b_\al\}=\{t=\td{t}^\al b_\al\}$. Now take a curve $C:[0,1]\ra \U,~\tau\mapsto x(\tau)$ on $\U$. Then, in $\V$, the tangent vector $T$ to $C$ at $x$ is
\bea
&& T={d b(x(\tau))\over d\tau}=b_\al~\dot{x}^\al\in T_x(\U),~~\dot{x}^\al={d x^\al(\tau)\over d\tau}.
\eea
However, the second derivative ${d^2b\over d\tau^2}={dT\over d\tau}$ is not tangent to $\U$. Thus we need to define a parallel ( projected or intrinsic) derivative $\nabla\over d\tau$ if we assume that for any vector $V$, only the component of $V$ that points along $\U$ is of interest to us such as would be the case of we live exclusively in $\U$. With the projector $P=b_\al (b_\al\cdot b_\beta)^{-1}b_\beta=b_\al g_{\al\beta}^{-1}b_\beta =b_\al g^{\al\beta}b_\beta$,
\bea
&&P\cdot P=P,~~P\cdot b_\al=b_\al\cdot P=b_\al,\nn\\
&&~~\Ra~~P\cdot t=t~~\forall t\in T_x(\U),
\eea
we can define the projected differential $\nabla=P\circ d$ as follows.

\bea
&&\nabla: \A\ra \A,~a\mapsto \nabla a=P\cdot da=b_\al g^{\al\beta}b_\beta\cdot da\nn\\
&&~~\Ra~~\nabla b_\al=b_\beta~\Gamma^\beta{}_\al,\nn\\
\label{Christoffel1}&&\Gamma^{\al}{}_{\rho}=g^{\al\beta}~b_\beta\cdot db_{\rho}= g^{\al\beta}~b_\beta\cdot b_{\rho'\rho}~dx^{\rho'}=\Gamma^{\al}{}_{\rho'\rho}~dx^{\rho'},~~b_{\rho'\rho}={\del^2_{\rho'\rho}}b,\\
&&\nabla b=P\cdot db=b_\al g^{\al\beta}~b_\beta\cdot db=b_\al g^{\al\beta}~b_\beta\cdot b_\al~dx^\al=b_\al~dx^\al\eqv db.\nn\\
&&\nabla^2 b=\nabla T ~dt=\nabla (b_\al dx^\al)= P\cdot d(b_\al dx^\al)=b_\al ~D dx^\al,\nn\\
&&D dx^\al=\Gamma^{\al}{}_{\rho}~ dx^\rho+ d^2x^\al.
\eea
Assuming that the state $\phi=const$ in $\U$ and with $a_1\cdot a_2=a_2\cdot a_1$, the Christoffel symbol $\Gamma$ defined in (\ref{Christoffel1}) can be written as
\bea
&&\Gamma^{\al}{}_{\rho'\rho}:=g^{\al\beta}~b_\beta\cdot b_{\rho'\rho}={1\over 2}~ g^{\al\beta}~(\del_{\rho'}g_{\beta\rho}-\del_\beta g_{\rho'\rho}+\del_\rho g_{\rho'\beta}).
\eea

[~Aside: It may also be useful to note that for any $T\in T(\U)=\bigcup_{z\in \U}T_z(\U)$,
\bea
\nabla T=P\cdot dT=d(P\cdot T)-dP\cdot T=dT-dP\cdot T
\eea
 This explicitly shows the role of the projector $P$ in determining the difference between the derivative $d$ and the projected derivative $\nabla$.~]

\subsection{Projected differential $\nabla$ on products}
Because of the projector $P$ in $\nabla=P\circ d$, $\nabla$ does not satisfy the Leibnitz rule with the algebra product $\star$ in $\A$, which is inherited by both the parametrized subset $\V$ and the tangent space $T_x(\U)$. That is,
\bea
&&\forall t,t_1,t_2\in T_x(\U),~~P\cdot t=t,\nn\\
&&P\cdot (t_1t_2)=\td{t}^\al\td{t}^\beta P\cdot(b_\al b_\beta)\neq t_1t_2\nn\\
&&~~\Ra~~\nabla(t_1t_2)\neq \nabla t_1~t_2+t_1\nabla t_2.
\eea

We will next try applying $\nabla$ on tensor products.

\subsubsection{Tensor algebra}

We will collect all tangent spaces and cotangent spaces (spaces of linear functionals on the tangent spaces) together to give the tangent and cotangent bundles $T(\U)=\bigcup_{p\in\U}T_p(\U),~~T^\ast(\U)=\bigcup_{p\in\U}T^\ast_p(\U)$ respectively.
\bea
&&T^\ast(\U)=\txt{Span}\{b^\al\},~~b^\al(b_\beta)=b^\al\cdot b_\beta~:=~\delta^\al_{\beta}\nn\\
&&~~\Ra~~b^\al\in ~g^{\al\beta}b_\beta+\ker~b_\al=(b_\al\cdot b_\beta)^{-1}b_\beta +\ker~b_\al.\nn\\
\eea

We consider the tensor algebras $\T(\mathcal{A}),\T({T(\U)})$ of $\mathcal{A},T(\U)$ with corresponding dual tensor algebras $\T({\mathcal{A}^\ast})\eqv \T^\ast({\A}),~\T({T^\ast(\U)})\eqv \T^\ast({T(\U)})$ for $\A^\ast,T^\ast(\U)$. Then $\nabla$ will satisfy the Liebnitz rule on a tensor since the projector $P$ will inherit a tensor product structure from $\T({\A})$.

We define the tensor algebra $\T(S)$ of any set $S=\txt{Span}\{s_\al\}=\{s=\td{s}^\al s_\al\}$ as follows.

\bea
&&\{s_\al\}^{\otimes n}=\{s_\al\}\otimes\{s_\al\}^{\otimes n-1}=\{s_{\al_1...\al_n}=s_{\al_1}\otimes s_{\al_2}\otimes...\otimes s_{\al_n}\} \nn\\
&& S^{\otimes k}=\txt{Span}(\{s_\al\}^{\otimes k})= \{\td{s}^{\al_1...\al_k}~s_{\al_1...\al_k}\} ,\nn\\
&&\T(S)=\bigoplus_{k=0}^\infty~S^{\otimes k}=\mathbb{C}\oplus S\oplus S^{\otimes 2}\oplus...\nn\\
\eea

Then the action of $\nabla$ becomes graded through the projector as follows.
\bea
&&\nabla T=P^{\otimes~ R(T)}\cdot dT,~~~~P^{\otimes R(T)}=P\otimes P^{\otimes~R(T)-1},\nn\\
&&\nabla (T_1\otimes T_2)=(P^{\otimes R(T_1)}\otimes P^{\otimes R(T_2)})\cdot d(T_1\otimes T_2)=(P^{\otimes R(T_1)}\otimes P^{\otimes R(T_2)})\cdot (dT_1\otimes T_2+T_1\otimes dT_2)\nn\\
&&~~~~=P^{\otimes R(T_1)}\cdot dT_1\otimes P^{\otimes R(T_2)}\cdot T_2+P^{\otimes R(T_1)}\cdot T_1\otimes P^{\otimes R(T_2)}\cdot dT_2\nn\\
&&~~~~=\nabla T_1\otimes T_2+T_1\otimes \nabla T_2,~~~~ T_1\in T(\U)^{\otimes R(T_1)},T_2\in T(\U)^{\otimes R(T_2)},\nn\\
\eea
where $R(T)$ refers to the tensorial rank of $T\in T(\U)^{\otimes R(T)}$.

The tensor product $\otimes$ and direct sum $\oplus$ have the same meaning as those for matrices; for example  a tensored positive linear functional $\Phi\in \T^\ast({\A})\eqv \T({\A^\ast})$ acts on the tensor algebra $\T_{\A}$ as follows.
\bea
\label{tensor-functional} &&\Phi:\T({\A})\ra \mathbb{C},~~t_1\otimes t_2\otimes...\mapsto\Phi(t_1\otimes t_2\otimes...)=\sum\phi_1(t_1)~\phi_2(t_2)~...\\
&&~~~~\eqv \sum(\phi_1\otimes \phi_2\otimes...)(t_1\otimes t_2\otimes...).\nn\\
\eea
That is, $\Phi$ behaves like the trace $\txt{Tr}$ over tensor products of matrices.
We model the action of the cotangent spaces $T^\ast(\U),~\T({T^\ast(\U)})\eqv  \T^\ast({T(\U)})$ on the tangent spaces $T(\U),~\T({T(\U)})$ along these lines.

\bea
&&\Phi(t)=\sum\td{\Phi}_{\al_1\al_2...}(b^{\al_1}\otimes b^{\al_2}\otimes...)~(\td{t}^{\beta_1\beta_2...}b_{\beta_1}\otimes b_{\beta_2}\otimes...)\nn\\
&&~~~~=\sum\td{\Phi}_{\al_1\al_2...}\td{t}^{\beta_1\beta_2...}(b^{\al_1}\otimes b^{\al_2}\otimes...)(b_{\beta_1}\otimes b_{\beta_2}\otimes...)\nn\\
&&~~~~=\sum\td{\Phi}_{\al_1\al_2...}\td{t}^{\beta_1\beta_2...}~b^{\al_1}(b_{\beta_1})b^{\al_2}(b_{\beta_2})...\nn\\
&&~~~~=\sum\td{\Phi}_{\al_1\al_2...}\td{t}^{\beta_1\beta_2...}\delta^{\al_1}_{\beta_1}\delta^{\al_2}_{\beta_2}...=\sum\td{\Phi}_{\al_1\al_2...}\td{b}^{\al_1\al_2...}
\eea

We can also introduce a metric tensor $g$ whose components are
\bea
&& g_{\al\beta}=b_\al\cdot b_\beta,~~g=g_{\al\beta}~b^\al\otimes b^\beta,\nn\\
&&d(b^\al\cdot b_\beta)=0=db^\al\cdot b_\beta+b^\al\cdot db_\beta= (P\cdot d b^\al)\cdot b_\beta+b^\al\cdot (P\cdot d b_\beta),\nn\\
&&\nabla b^\al\cdot b_\beta=-b^\al\cdot \nabla b_\beta=-\Gamma^\rho{}_\beta~b^\al\cdot b_\rho=-\Gamma^\al{}_\beta=-\Gamma^\al{}_\rho~b^\rho\cdot b_\beta\nn\\
&&~~\Ra~~~~\nabla b^\al=-\Gamma^\al{}_\beta~b^\beta.
\eea

\subsection{Tensorial derivative $\del$}\label{tensoring-derivative}
It is possible to define the derivative in such a way that it increases the rank of the tensor by $1$. This derivative will satisfy a ``symmetrized'' form of the Leibnitz rule if the tensor product is ``symmetrized''. A totally symmetric tensor product will satisfy the usual Leibnitz rule. We assume that the tensor product is associative.

\bea
&&{\del}=b^\al\otimes \nabla_\al:~T^\ast(\U){}^{\otimes p}\ra T^\ast(\U){}^{\otimes~ 1+p},\nn\\
&&{\del}T=~b^\al\otimes\nabla_\al T=b^\al\otimes P^{\otimes R(T)}\cdot\del_\al T= P\cdot b^\al\otimes P^{\otimes R(T)}\cdot\del_\al T  ,\nn\\
&&~~~~=P^{\otimes~ 1+R(T)}\cdot(b^\al\otimes\del_\al T).\nn\\
&&{\del}(T_1\otimes T_2)=b^\al\otimes\nabla_\al(T_1\otimes T_2)=b^\al\otimes\nabla_\al T_1\otimes T_2+b^\al\otimes T_1\otimes \nabla_\al T_2\nn\\
&&~~~~={\del}T_1\otimes T_2+b^\al\otimes T_1\otimes \nabla_\al T_2.\nn\\
\eea

Therefore, for $V=b_\al\td{V}^\al$,
\bea
&&{\del}V=b^\al\otimes b_\beta~(\del_\al \td{V}^\beta+\Gamma^\beta{}_{\al\delta}\td{V}^\delta) =b^\al\otimes b_\beta~D_\al \td{V}^\beta,\nn\\
&&{\del}^2V=b^\al\otimes b^\beta\otimes b_{\gamma}~D_\al D_\beta \td{V}^\gamma,
\eea
and so on.
\subsection{Higher derivatives: Force, Curvature}
The first projected derivative of the tangent- (or velocity-) vector has the meaning of acceleration as in classical mechanics, meanwhile the second projected derivative of the velocity vector describes ``intrinsic'' curvature of $\U$.

``Straight'' lines or geodesics in $\U$ correspond to curves $x:[0,1]\ra \U,~\tau\ra x(\tau)$ such that
\bea
&&\Gamma^{\al}{}_{\rho}~ dx^\rho+ d^2x^\al=0,\nn\\
&&\ddot{x}^\al=-g^{\al\beta}~b_\beta\cdot \dot{b}_\rho~\dot{x}^\rho=f^\al{}_\rho~\dot{x}^\rho,\nn\\
&&\dot{x}^\al={dx^\al\over d\tau}={dt\over d\tau}{dx^\al\over dt}=\gamma ~{dx^\al\over dt},~~\gamma={c\over \sqrt{g_{\al\beta}{dx^\al\over dt}{dx^\beta\over dt}}},\nn\\
&&d\tau~=~{1\over c}~ds~=~{1\over c}~\sqrt{g_{\al\beta}{dx^\al}{dx^\beta}},
\eea
where $c$ is the speed of light in vacuum.

If $\phi$ is the Gibbs state $\omega: a\mapsto \omega(a)={\txt{Tr}(a~e^{-\beta H})\over \txt{Tr}~e^{-\beta H}},~~\beta={1\over kT}$ of statistical mechanics for example, then a particle in a thermal bath in equilibrium at temperature $T$  will experience a fictitious field of force (in unit of frequency $s^{-1}$)
\bea
&&f^\al{}_\beta=-~g^{\al\rho}~b_\rho\cdot \dot{b}_\beta=-~(\omega(b^\dg_\al b_\rho+  b^\dg_\rho b_\al))^{-1}~\omega (b^\dg_\rho \dot{b}_\beta+ \dot{b}^\dg_\beta b_\rho),\nn\\
&&\dot{b}_\al={d b_\al\over d\tau}={dt\over d\tau}{d b_\al\over dt} =\gamma~(~{\del b_\al\over\del t}+{i\over \hbar}[H,b_\al]~),~~\gamma={c\over \sqrt{g_{\al\beta}{dx^\al\over dt}{dx^\beta\over dt}}},
\eea
where $b_\al$ may be taken to be the quantum mechanical momentum and position operators or the creation and anihilation operators constructed from them, or simply phase space coordinates.

Strings and membranes in a thermal bath will equally experience similar kinds of forces.

\textbf{Remarks:}
\begin{itemize}
\item The set of real, symmetric, positive-determinant metrics of $g$-type $\{g\}$ on $\U$ is at least as large as the set of positive linear functional $\A_+^\ast$. The remainder of the linear functionals $\A^\ast\backslash\A^\ast_+$ also produce corresponding metrics of their own which may neither be real nor symmetric nor positive-determinant.
\item Torsion effects would be present if one includes the anti-symmetric piece ${1\over 2}\phi(a_1^\ast a_2-a_2^\ast a_1)$ of the dot product, so that $a_1\cdot a_2=\phi(a_1^\ast a_2)\neq a_2\cdot a_1=\phi(a_2^\ast a_1)$, since the metric would no longer be symmetric. The latter dot product is complex, implying a complex non-symmetric metric but we can consider a class of real non-symmetric dot-products: $a_1\cdot_\ld a_2=\phi(\ld a_1^\ast a_2+\bar{\ld}a_2^\ast a_1),~~\ld\in \mathbb{C}$. For applications of torsion, see references such as \cite{saa1993,moffat1994}.
\end{itemize}

\subsection{Quantum Fields in Space-time}
The (linear) geometry of space-time is determined by the (linear) functionals on the algebra(s) of the fields defined  on the space-time.

Suppose  $\V=\{A(x),~x\in \U\}$ is the value space of a given quantum field $A$ defined in space-time $\U$. If one then imagines the value space $\V$ as being a subspace of the covering space (or inclusive algebra) $\A$ of all the quantum field values of type $A(x)$, then we can imagine space-time $\U$ to be curved as a result of the quantum field $A:\U\ra \V \subseteq\A ,~x\ra A(x)$. The metric will be the dot-product of the derivatives of the quantum field $A$ and the derivative is the projected derivative $\nabla=P\circ d$. That is, $G=G_A=(g_{\al\beta})=(A_\al\cdot A_\beta)=(\del_\al A\cdot \del_\beta A)$,~~$P=P_A=A_\al g^{\al\beta}A_\beta$.

If two quantum fields $A,B$ are present in space-time $\U$, then one can consider the following two situations:

\subsubsection{$A$ is dynamical and $B$ is non-dynamical}

An external or applied (non-dynamical or condensed) quantum field $B$, as a background field, would affect the dynamics of a  quantum field $A$ by modifying the curvature of space-time through the metric $G=G_A\oplus G_B
$ and  projected derivative $\nabla=P\circ d$ with the projector $P=P_A+P_B$.

\subsubsection{Both $A$ and $B$ are dynamical}
In this case, we imagine the covering algebra $\C$ to be locally isomorphic to (written as $\simeq$) a direct product of the covering algebras $\A,\B$ of the fields considered separately. That is, $\C\simeq \A\times \B$. The value space $\V$ also has the local direct product structure $\V\simeq \V_A\times \V_B,$ where $\V_A=\{A(x),~x\in \U\}$ and $\V_B=\{B(x),~x\in \U\}$. The projected derivative is again given by $\nabla=P\circ d=(P_A+P_B)\circ d=\nabla_A+\nabla_B$.

\subsection{Some examples}
These examples are for illustrative purposes only and so need not be taken too seriously.
\begin{enumerate}

\item Let $\A=\mathbb{R}^n_q=(\mathbb{R}^n,\ast)=\{x=(x^1,x^2,...,x^n)\eqv\txt{diag}(x^1,x^2,...,x^n):\mathbb{R}^n\ra \mathbb{R}^n,~x^ix^j\eqv x^i\ast q^j=q^{ij}x^jx^i,~~q^{ji}={1\over q^{ij}}\}$ be a $q$-deformed version of the $\mathbb{R}^n$ considered in the preliminaries section (\ref{preliminaries}).

 Then ~ $\V=\V_{q(\theta)}=\{\vec{x}:\U\ra \mathbb{R}^n,~(x^ix^j)(u)\eqv(x^i\ast x^j)(u):=x^i(u)e^{{i\over 2}{\ola{\del}\over \del u^\al}\theta^{\al\beta}{\ora{\del}\over \del u^\beta}}x^j(u)\}$~ may be regarded as a realization for a parametrized subspace of~ $\mathbb{R}_q^n$, with $q=q(\theta)=e^{{i\over 2}\ola{\del}\wedge\ora{\del}}=e^{{i\over 2}{\ola{\del}\over \del u^\al}\theta^{\al\beta}{\ora{\del}\over \del u^\beta}}$. The parameter space $\U$ or hypersurface may be chosen to be isomorphic to the Moyal plane  ~$\M=\{\hat{u}=(\hat{u}^1,\hat{u}^2...,\hat{u}^p),~u^\al=\hat{u}^\al(u),~\hat{u}^\al\in \V_{q(\theta)},~\hat{u}^\al \ast\hat{u}^\beta-\hat{u}^\beta \ast \hat{u}^\al=i\theta^{\al\beta}=-i\theta^{\beta\al}~\}$. That is,
 \bea
 &&\U\simeq\{u=(u^1,u^2...,u^p),~u^\al=\hat{u}^\al(u),~\hat{u}\in \M~\}.
 \eea

 In the case where the $\theta^{\al\beta}$'s are constant, the set $B=\{\vec{x}_\al={\del\vec{x}\over\del u^\al}\}$ is a basis for the tangent space since partial derivatives commute.

The metric is
\bea
&&G=(g_{\al\beta})=(\vec{x}_\al\cdot\vec{x}_\beta)={1\over 2n}\sum_i(x^i_\al\ast x^i_\beta+x^i_\beta\ast x^i_\al),~~~~\vec{x}_\al=\del_\al\vec{x}={\del\vec{x}\over\del u^\al}.\nn\\
&&P=g^{\al\beta}\vec{x}_\al\otimes \vec{x}_\beta\eqv\vec{x}_\al g^{\al\beta}\vec{x}_\beta\eqv (x^i_\al g^{\al\beta}x^j_\beta).
\eea
The rest of the analysis continues in a straightforward manner. For a review of quantum fields on the Moyal plane, see \cite{qft-us} for example.

\item We may take $\A=\{a=ad_X:\pounds\ra\pounds,~~X\in \pounds\}=ad_\pounds$~ to be the adjoint representation of some Lie algebra $\pounds$. $ad_{X}(Y)=[X,Y],~~a^\ast=a^\dagger$, where $d$ is the dimension of the representation. We then choose $\phi(a)={1\over d}\txt{Tr}~a={1\over d}\txt{Tr}~ad_X$.
    Therefore,~ $a\cdot b={1\over 2}~{1\over d}\txt{Tr}~(a^\ast b+b^\ast a)={1\over 2}~{1\over d}\txt{Tr}~(ad_{X}~^\ast\circ ad_{Y}+ad_{Y}~^\ast\circ ad_{X})=\mathbb{I}~g(X,Y)$ would be the Cartan-Killing form on $\pounds$.

\item If $\A=GL(n,\mathbb{C})$, the set of all complex invertible $n\times n$ matrices (regarded as invertible linear maps from $\mathbb{C}^n$ to $\mathbb{C}^n$), then a parametrized subset $\V$ could be a representation of any of the continuous (Lie) groups such as $SL(n,\mathbb{C}), O(n,\mathbb{C}),~ U(n,\mathbb{C}),$ etc. $\U=\{\ld\in \mathbb{R}^{\dim(\V)}\}$ would be the parameter space of the Lie group. $a^\ast=a^\dagger,~~ \phi(a)={1\over \dim(\V)}\txt{Tr}~a.$
For example, let $=SL(n,\mathbb{C})=\{g(\ld),~~\det g=1\}$ be generated by the basis elements $\{J_\al=J(\ld)={\del \ln g\over\del\ld^\al}\eqv g^{-1}\del_\al g,~~\txt{Tr}J_\al={\del_\al\det g\over\det g}=0,~~\al=1,2,..., 2n^2-2)\}$. Then since $\txt{Tr}[J_\al,J_\beta]=0$, we may be able to write $[J_\al,J_\beta]\eqv-(\del_{\al}J_{\beta}-\del_{\beta}J_{\al})=f_{\al\beta}^\rho(\ld)~J_\rho$ ~up to transformations of the form $J_\rho\ra h^{-1}J_\rho h.$

Choose $\V=\Ad_{ SL(n,\mathbb{C})}=\{\Ad_g,~g\in SL(n,\mathbb{C}),~\Ad_g:SL(n,\mathbb{C})\ra SL(n,\mathbb{C}),~h\mapsto g^{-1}hg\}$, adjoint representation of $SL(n,\mathbb{C})$, then we can use the projector
\bea
&&P=\ad_{J_\al} g^{\al\beta} \ad_{J_\beta},~~~\ad_XY=[X,Y],\nn\\
&&g_{\al\beta}={1\over 2}~{1\over 2n^2-2}\txt{Tr}~(\ad_{J_\al}{}^\ast~\ad_{J_\beta}+\ad_{J_\beta}{}^\ast~ \ad_{J_\al})\nn\\
&&~~~~={1\over 2}~{1\over 2n^2-2}~(\bar{f}_{\al\rho}^\gamma f_{\gamma\beta}^\rho+\bar{f}_{\beta\rho}^\gamma f_{\gamma\al}^\rho)
\eea
to construct the projected derivative on $\U$,~\\ $T_\ld(\U)=span(B)=span\{b_\al=\ad_{J_\al},~[b_\al,b_\beta]=f_{\al\beta}^\rho~ b_\rho\}=\{X\}$,~~~~ $[ad_X,\ad_Y]=\ad_{[X,Y]}$.
\bea
\nabla X=P\cdot dX
\eea

\item \emph{Hamiltonian system}:
Let $\A=\{a:[0,1]\times\mathbb{R}^{2n}\ra \mathbb{C}\}\eqv \{a=(a_t,t\in[0,1]),~~a_t:\mathbb{R}^{2n}\ra \mathbb{C}\}$ be the algebra of complex functions on $[0,1]\times\mathbb{R}^{2n}$.
Then a Hamiltonian system $\V$ in $\A$ with Hamiltonian $H\in \A$ is defined by giving the following differentiation rule.

For any map $f:[0,1]\times\D\ra \mathbb{C},~~(t,p,q)\ra f(t,p,q)$,
\bea
&&{df\over dt}={\del f\over\del t}+\{H,f\}_{pb},~~~~\{a,bc\}_{pb}=\{a,b\}_{pb}c+b\{a,c\}_{pb}.
\eea
Therefore, unless the Hamiltonian $H$ is restricted in some way, there will be at least as many Hamiltonian systems as there are elements of $\A$ since every element of $\A$ can be considered as a Hamiltonian.
The parameter space here may be chosen to be $\U=\{p\in \mathbb{R}^n\}\simeq\{q\in \mathbb{R}^n\}$.
We may also define the state
\bea
&&\phi(f)\eqv F[q]=\int_{t\in[0,1]} dt ~f(t,p(t),q(t)),~~~~f^\ast=\bar{f}.
\eea

The classical case $\{a,b\}_{pb}={\del a\over\del p}{\del b\over\del q}-{\del b\over\del p}{\del a\over\del q}$ may be extended as follows: replace the canonical variables $x=(p,q)$ with $x_\ast=(p_\ast,q_\ast)$, where
\bea
&& (p,q)\ra (p_\ast ,q_\ast)=(p e^{{1\over 2}({\ola{\del}\over\del p}{\ora{\del}\over\del q}-{\ola{\del}\over\del q}{\ora{\del}\over\del p}) },q e^{{1\over 2}({\ola{\del}\over\del p}{\ora{\del}\over\del q}-{\ola{\del}\over\del q}{\ora{\del}\over\del p}) })\eqv (p e^{{1\over 2}\ola{\del}\wedge\ora{\del} },q e^{{1\over 2}\ola{\del}\wedge\ora{\del} }).\nn\\
\eea
Then
\bea
&&f(t,p,q)\ra f_\ast=f(t,p_\ast,q_\ast):=\int d^n\td{p}d^n\td{q}~\td{f}(t,\td{p},\td{q})~e^{i(\td{p}p_\ast+\td{q}q_\ast)}~~(\txt{Weyl
ordering choice}),\nn\\
&&\dot{f}_\ast={df_\ast\over dt}={d\over dt}f(t,p_\ast(t), q_\ast(t)):={\del f_\ast\over\del t}+[H_\ast,f_\ast]={\del f_\ast\over\del t}+H_\ast f_\ast-f_\ast H_\ast,\nn\\
&&{d(f_\ast g_\ast)\over dt}=[H_\ast,f_\ast g_\ast]=[H_\ast,f_\ast]~ g_\ast+f_\ast~[H_\ast, g_\ast]={df_\ast\over dt}~g_\ast+f_\ast {dg_\ast\over dt}.
\eea
The $\ast$ operation is a particular case of $\pi$ in the equation $\pi(f)\pi(g)=\pi(f\ast g)$, where $(f\ast g)(x)=f(x)e^{{1\over 2}\ola{\del}\wedge\ora{\del} }g(x)$.

In the undeformed case, the tangent space is spanned by $\{b_i=\dot{q}_i=\{H,q_i\}_{PB}={\del H\over\del p_i}\}$ and so the projector unto the tangent space is
\bea
&&P=b_i(b_i\cdot b_j)^{-1}b_j.
\eea

Similarly, in the deformed case, the tangent space is spanned by $\{b_i=\dot{q}_{\ast i}=[H_\ast,q_{\ast i}]\}$ and one also has a projector $P$ onto the tangent space.

The projector $P$ can then be used to define a projected differential ~$\nabla=P\circ d$~ as before.

\item $\A=\{\Phi:\G\ra \mathbb{C},~\theta\mapsto \Phi(\theta)\}$ could be the algebra of functions of Grassman variables $\G=\{\theta\}$ meanwhile $\V=\{\Phi_x\}$ would be the value set(s) of (reducible or irreducible) superfield(s) $\Phi: \mathbb{R}^n\ra \A,~x\ra \Phi_x$. The operation $^\ast$ is the usual complex (or hermitian) conjugation $^\dg$. We take Grassman integrals (integration over Grassman variables) to be linear functionals $\phi$'s on $\A$. The parameter space would then be $\U=\{x\}=\M^n$. That is $\phi(\Phi)=\int d\theta~\Phi_x(\theta)$. One can also investigate effects of non-commutativity in supersymmetry such as in \cite{babar2007}.

\item $\A$ could be the space of all random variables. $\V\subset \A$ be the value space of quantizable fields $\{\vphi\}$ of any classical field theory whose dynamics is based on an action principle with an action $S=S[\{\vphi\}]$. $\U$ would be space-time, $\phi$ would be a quantum measure such as the path integral measure and $a^\ast=a^\dagger$ :
\bea
&&\phi(a)={1\over \N}\int D\{\vphi\}~a~e^{{i\over\hbar}S[\{\vphi\}]},~~~~\N=\int D\{\vphi\}~e^{{i\over\hbar}S[\{\vphi\}]}.\nn\\
\eea
We may as usual take the components of the partial derivatives $\{\del_\mu\vphi\}$ (which are parallel to the momentum of the fields as a Noether current corresponding to translational invariance) as a basis for the tangent space $T(\U)$. However, the components of any of the ``conserved''(or Noether) currents (which are associated with the symmetries of the action $S[\{\vphi\}]$~) may be used as a basis for a physically realizable tangent space $T_{ph}(\U)$. Here, $f(\{\vphi\})={1\over\N}~e^{{i\over\hbar}S[\{\vphi\}]}$~ may be regarded as a joint probability density function (jpdf) for the random variables $\{\vphi\}$, and so one can define conditional probability density functions (cpdf's) and hence conditional states corresponding to the cpdf's.

A similar procedure can be followed for other statistical field theories as well.

\item We can also consider the parametrization (parameters include rest mass $m$, electric charge $e$, speed of light in vacuum $c$, the reduced Plank constant $\hbar$, Newton's gravitational constant $G$, Boltzmann constant $k=k_B$ and so on) and or dynamics of universes \emph{based} on or at our own universe.

We may interpret the metric $g$ as a means of communication between vectors pointing in different directions, concerning how the vectors need to fit together in order to generate variously skewed frames of various sizes on which various shapes can then be constructed.

Let $\A=\{\vphi:x\in \M\ra \vphi(x)\}$ be the space of all quantum fields that can be localized in space-time $\M$. The value space of physically realizable quantum fields is given by $\V=\{\vphi_x(u),~u=(u^1,u^2,...):=(m,e,c,\hbar,G,k,...),~x\in D\subset \M\}$,~~ where the parameter space $\U=\{
u\}$ is the set of all possible values that the physical constants can assume in various universes.
We will take a basis for parallel displacement along the parameter space $\U$ to be $B=\{\vphi_I\eqv\vphi_{\al x}={\del\vphi_x\over\del u^\al},~x\in D,~\al=1,2,...\}$. The quantum field ~$\vphi:x\in \D\subset \M\ra \vphi(x)$~ is supported in the finite region $D$ of space-time.

If gauge invariance is desired, eg. for a universe in a background gauge field with potential $A$, then we may extend the operation $^\ast$ thus:
\bea
&&\vphi^\ast_x=\vphi^\dg_x~\P e^{ie\int^{\hat{x}}_x A},~~~~(\hat{x}^\mu\vphi)_x=x^\mu\vphi_x,\nn\\
&&\phi(\vphi_x)=\langle0|T(\vphi_x)|0\rangle,\nn\\
\eea
where $\P,T$ denote path ordering and time-ordering respectively, $\hat{x}$ is a position operator, $A$ is differential 1-form gauge field and $\phi$ is the vacuum state of the quantum field theory. The metric
\bea
&& g_{IJ}\eqv g_{\al x~\beta y}=\phi(\vphi^\ast_{\al x}\vphi_{\beta y})=\langle 0|T(\vphi^\dg_{x\al} \P e^{ie\int^{y}_x A}\vphi_{\beta y})|0\rangle\eqv \vphi_I\cdot\vphi_J,
\eea
is therefore a causal and gauge-invariant ``propagator''. The corresponding projector can then be written down; $P=\vphi_Ig^{IJ}\vphi_J=\vphi_Ig^{-1}_{IJ}\vphi_J\eqv \vphi_I(\vphi_I\cdot\vphi_J)^{-1}\vphi_J $.

\end{enumerate}

\subsection{Analytic continuation and Einstein gravity}\label{ACEG}

Analytic continuation, introduced for the dynamical aspect of space-time, is achieved through a Lagrange multiplier. The differential $\hat{d}$ used in this section \ref{ACEG}  is the exterior differential and the product is the exterior product.

Consider space-time $\U\simeq \{x\}$ and represent all the fields in some region $\D\subset\U$ by a grand field $A$ and denote its local value space as $\V \simeq \{A(x)\}$. $\A\supseteq \V$ is then the grand algebra, of all fields supported in the region $\D\subset\U$. The linear functionals on $\A$ should contribute in determining the geometry of $\D$ and hence that of $\U$ by analytic continuation.

We will denote the local metric, connection and curvature due to the grand field $A$ by $g^A$, $\omega$ and $\Omega$ respectively. We also denote the analytically continued local metric, connection and curvature by $g$, $\Gamma$ and $R$ respectively.

Then the curvature $2$-forms $\Omega^\al{}_\beta$ and $R^\al{}_\beta$ satisfy the 2nd Bianchi identity simultaneously in $\D\subset \U$.
\bea
\label{ACEG-Bianchi1}&& \hat{\nabla}_A\omega^\al{}_\beta= \hat{d}\omega^\al{}_\beta+\omega^\al{}_\rho \Omega^\rho{}_\beta-\omega^\rho{}_\beta \Omega^\al{}_\rho =0\\
&&~~~~=(\bar{\P} e^{-\int_{\gamma} \omega}~\hat{d}~ \P e^{\int_{\gamma} \omega}) \Omega,\nn\\
\label{ACEG-Bianchi2}&& \hat{\nabla} R^\al{}_\beta=\hat{d}R^\al{}_\beta+\Gamma^\al{}_\rho R^\rho{}_\beta-\Gamma^\rho{}_\beta R^\al{}_\rho =0\nn\\
&&~~~~=(\bar{\P} e^{-\int_{\gamma} \Gamma}~\hat{d}~ \P e^{\int_{\gamma} \Gamma}) R,
\eea
where $\P$ and $\bar{\P}$ denote path ordering and reverse path ordering (Appendix \ref{Path-ordering}) respectively.
Therefore in $\U$,~ $\Omega$ and $R$ may be related through a Lagrange multiplier $\ld$.

\bea
\label{EinsteinEqns}&&\td{R}^\al{}_\beta =\ld^{\al\mu}{}_{\beta\nu}~\td{\Omega}^\nu{}_{\mu}+C^\al{}_\beta,\\
&&\td{R}^\al{}_\beta=(\P e^{\int_{\gamma} \Gamma}~R)^\al{}_\beta=(\P e^{\int_{\gamma} \Gamma})^\al{}_\mu~(\bar{\P} e^{-\int_{\gamma} \Gamma})^\nu{}_\beta~R^\mu{}_\nu,\nn\\
&&\td{\Omega}^\al{}_\beta=(\P e^{\int_{\gamma} \omega}~R)^\al{}_\beta=(\P e^{\int_{\gamma} \omega})^\al{}_\mu~(\bar{\P} e^{-\int_{\gamma} \omega})^\nu{}_\beta~\Omega^\mu{}_\nu,
\eea
where $\ld$ is a $4$-array of closed $0$-forms (constant functions)  and $C$ is a $2$-array of closed $2$-forms ($\hat{d}\ld=0,~\hat{d} C=0$). The major point here is that meanwhile $\td{\Omega}$ has support only in $\D\subset \U$, ~~$\td{R}$ and $C$ can be supported anywhere in $\U$.

(\ref{EinsteinEqns}) may be considered as an analog of Einstein's equations for the gravitational field.

\section{Conclusion}\label{conclusion}
We have reviewed linear projection geometry on a $C^\ast$-algebra $\A$ using the Cauchy-Schwartz inequality for positive linear functionals on $\A$ as a natural tool for that purpose. The physics of statistical and quantum measurements was mentioned. We also reviewed calculus on hypersurfaces in $\A$, where linear projection geometry played an important role.
Finally, we discussed how a field supported in a certain region $\D$ of space-time $\U$ can produce curvature in the support $\D$ and that this curvature could then be analytically continued to all of $\U$ with the help of the 2nd Bianchi identity. This analytic continuation was intended to make contact with Einstein's field equations for gravity.

\begin{center}
\textbf{Acknowledgements}
\end{center}
This work was supported by the US Department of Energy under grant number DE-FG02-85ER40231.
The author is grateful to Prof. A. P. Balachandran for the ideas that made this work possible.

\appendix

\section{In an orthonormal frame}\label{orthon-frame}
We wish to see how the connection $1$-form $\Gamma$ and curvature 2-form $R$ look in an orthonormal basis for the tangent space.

Assume that we have a local orthonormal frame (a basis $\{b_\al\}$ can be orthonormalized, section \ref{Ortho}, to give a basis $\{\hat{b}_\al\}$ such that $g_{\al\beta}=\hat{b}_\al\cdot \hat{b}_\beta=\delta_{\al\beta}$). Then since
\bea
&&P=b_\al g^{\al\beta}b_\beta,\nn\\
&&dg_{\mu\nu}=db_\al\cdot b_\beta+b_\al\cdot db_\beta\nn\\
&&~~~~=db_\al\cdot b_\beta+b_\al\cdot db_\beta=\nabla b_\al\cdot b_\beta+b_\al\cdot \nabla b_\beta=\Gamma^{\rho}{}_\al~b_\rho\cdot b_\beta+\Gamma^\rho{}_\beta b_\al\cdot b_\rho\nn\\
&&~~~~=\Gamma^{\rho}{}_\al~g_{\rho\beta}+\Gamma^\rho{}_\beta g_{\al\rho}=\Gamma_{\beta\al}+\Gamma_{\al\beta}=b_{\beta}\cdot db_{\al}+b_{\al}\cdot db_{\beta},\nn\\
&&R_{\al\beta}=g_{\al\rho}R^\rho{}_\beta=g_{\al\rho}(d\Gamma^\rho{}_\beta+\Gamma^\rho{}_\gamma\Gamma^\gamma{}_\beta) ,
\eea
we have
\bea
&&\hat{P}= \hat{b}_\al \delta^{\al\beta}\hat{b}_\beta=\hat{b}_\al \hat{b}_\al,\nn\\
&&\hat{\Gamma}^\al{}_\beta=\hat{\Gamma}_{\al\beta}=-\hat{\Gamma}_{\beta\al}=\hat{b}_\al\cdot d\hat{b}_\beta,\nn\\
&&\hat{R}^\al{}_\beta=\hat{R}_{\al\beta}=d\hat{\Gamma}_{\al\beta}+\hat{\Gamma}_{\al\gamma}\hat{\Gamma}_{\gamma\beta}=-\hat{R}_{\al\beta}.
\eea
Therefore, in $2$ dimensions in particular, $R$ is exact in an orthonormal basis and the lone non-zero component of $R$ (respectively $\Gamma$)  is $\hat{R}_{12}$ (respectively $\hat{\Gamma}_{12}$)~;
\bea
\hat{R}_{12}=d\hat{\Gamma}_{12}=d\hat{b}_1\cdot d\hat{b}_2=({\del \hat{b}_1\over\del u^1}\cdot {\del \hat{b}_2\over\del u^2}-{\del \hat{b}_1\over\del u^2}\cdot {\del \hat{b}_2\over\del u^1})du^1 du^2 .
\eea
\section{Path ordering}\label{Path-ordering}
Consider a $p$-form $f$ and a $1$-form $A$ on a space $\mathcal{D}$ related by the following first order differential equation.
\bea
&&\hat{d}f=Af.\nn\\
\eea
Then we have a solution of the form
\bea
&&f=f(x,{\gamma}_x)=\mathcal{P} e^{\int_{\gamma_x} A}C(x),
\eea
where $\gamma=\gamma_x: [\tau_0,\tau]\subset\mathbb{R}\ra \mathcal{D},~s\ra \gamma(s),~\gamma(\tau)=x$, is any curve that ends at $x$, $\mathcal{P}$ denotes path ordering and $C$ is a closed $p$-form, $\hat{d}C=0$.

\bea
&&\mathcal{P} e^{\int_{\gamma_x} A}=1+\sum_{k=1}^\infty~\int_{[\tau_0,\tau]^k}~{1\over k!}~ \mathcal{P}(A_1A_2...A_k),
\nn\\
&&\mathcal{P}(A_1A_2...A_k)=A^\theta_{(12...k)}\eqv \sum_{\sigma\in S_k}A^\theta_{\sigma(1)\sigma(2)...\sigma(k)},\nn\\
&&A^\theta_{12...k}=\theta_{12...k}A_1A_2...A_k,~~A_i\eqv A(\gamma(s_i))=A_\al(\gamma(s_i))~{\hat{d}\gamma^\al(s_i)},~i=0,1,...,k,\nn\\
&&\theta_{12...k}=\prod_{i<j} \theta(s_i-s_j)=\theta(s_1-s_2)\theta(s_1-s_3)..\theta(s_1-s_k)\theta(s_2-s_3)..\theta(s_2-s_k)...\theta(s_{k-1}-s_k),\nn\\
\eea
 where $S_k$ is the permutation group and $\theta$ is the Heaviside function.

Also, one may write the path ordered integral as a product integral \cite{karp-mansouri-1999},
\bea
&&\mathcal{P} e^{\int_{\gamma_x} A}=\prod_{s=\tau_0}^{\tau}e^{A(\hat{d}s)}:=\lim_{\Delta s\ra 0}(\prod_{k=1}^\infty e^{A(\hat{\Delta} s_k)})|_{s_1=\tau_0,~s_\infty=\tau},\nn\\
&&A(\hat{d}s)=A_i(\gamma(s))~{d \gamma^i(s)\over ds}ds,~~~~\sum_k\Delta s_k=\tau-\tau_0.
\eea

When $\gamma$ is a closed curve $\gamma(\tau)=\gamma(\tau_0)=x$, then with the help of Stokes' theorem,

\bea
&&\mathcal{P} e^{\oint_{\gamma_x} A}=\prod_{Int(\gamma)}e^{DA (\hat{d}u^1,\hat{d}u^2)}= \prod_{u^a=(u^1,u^2)\in [\tau^0,\tau]^2 }e^{DA (\hat{d}u^1,\hat{d}u^2)|_{Int(\gamma)}},~~DA=\hat{d}A+A^2\eqv F_{ij}\hat{d}x^i\hat{d}x^j,\nn\\
&&DA (\hat{d}u^1,\hat{d}u^2)=F_{ij}{\del (x^i,x^j)\over \del (u^1,u^2)}du^1du^2=F_{12}du^1du^2,\nn\\
\eea
where  $Int(\gamma)$ is any interior (see appendix \ref{PStopology}) in $\mathcal{D}$ of the closed curve $\gamma$.

\section{Point Set topology}\label{PStopology}
A topological space $T_S=\{o,~o\subseteq S\}$ over a set $S$ is a subset of the power set (set of all subsets) of $S$ that is invariant under union $\cup$ and intersection $\cap$.
\bea
\cup,\cap:T^n_S=T_S\times T^{n-1}_S \ra T_S,~(o_1,o_2,...,o_n)\mapsto \cup_io_i,~ \cap_io_i.
\eea
$T_S$ is an algebra where union $\cup$ and intersection $\cap$ play the role(s) of addition $+$ and multiplication $\ast$ respectively.

The elements of $T_S$ are referred to as open sets. A closed set is the complement $C(o)$ of some $o\in T_S$ defined thus:
\bea
o\cup C(o)=S,~~o \cap C(o)=\emptyset\eqv \{~\}.
\eea
Let's denote the complementary or dual (or closed) topology by $ C(T_S)=\{C(o),~o\in T_S\}$.
The interior $A^o\eqv Int(A)$ of any $A\subset S$ is the union $\cup$ of all open subsets of $A$ and the closure $\bar{A}\eqv Cls(A)$ of any $A\subset S$ is the intersection $\cap$ of all closed supersets of $A$. The boundary $B(A)$ of $A$ may be defined implicitly by
\bea
\bar{A}=A^0\cup B(A).
\eea
$A$ is open iff $A=A^0$ and $A$ is closed iff $A=\bar{A}$.

\newpage
\bibliographystyle{apsrmp}

 \ed